\pdfoutput=1
\documentclass{article}
\usepackage[a4paper]{geometry}
\usepackage{graphicx}
\usepackage{amssymb,amsmath,amsthm}
\usepackage{epstopdf}
\usepackage{tikz}
\usepackage{sidecap}
\usepackage{subcaption}
\usepackage{cite}
\usepackage{authblk}
\usepackage{appendix}
\usepackage[labelfont={bf,small},textfont={small}]{caption}

\theoremstyle{definition}
\newtheorem{definition}{Definition}[section]

\renewcommand{\vec}[1]{{\boldsymbol{#1}}}
\renewcommand{\epsilon}{\varepsilon}
\DeclareMathOperator\sgn{sgn}

\title{Parabolic Metamaterials and Dirac Bridges}
\author[1]{D. J. Colquitt}
\author[2]{N. V. Movchan}
\author[2]{A. B. Movchan}
\affil[1]{Department of Mathematics, Imperial College London, London, SW7 2AZ, UK.}
\affil[2]{Department of Mathematical Sciences, University of Liverpool, Liverpool L69 3BX, UK.}

\date{\today}                       

\begin{document}

\maketitle

\begin{abstract}
A new class of multi-scale structures, referred to as  `parabolic metamaterials' is introduced and studied in this paper.
For an elastic two-dimensional triangular lattice, we identify dynamic regimes, which corresponds to so-called `Dirac Bridges' on the dispersion surfaces.
Such regimes lead to a highly localised and focussed unidirectional beam when the lattice is excited.
We also show that the flexural rigidities of elastic ligaments are essential in establishing the `parabolic metamaterial' regimes.  
\end{abstract}

\section{Introduction}

Metamaterials are synthetic materials that are designed to exhibit surprising properties that are rarely found in nature.
Following the first practical demonstration of negative refraction~\cite{shelby2001experimental} and the perfect lens~\cite{grbic2003subwavelength, fang2005sub,melville2005super}, there has been an explosion of interest in metamaterials.
One of the most unusual classes of metamaterial structures is the class of hyperbolic metamaterials, which currently enjoys significant interest in optics and plasmonics~\cite{poddubny2013hyperbolic}, as well as acoustics and solid mechanics~\cite{langley1996response,langley1997response,Ruzzene2003,ayzenberg2008resonant,osharovich2010wave,colquitt2012dynamic,movchan2014resonant}.
Optical hyperbolic metamaterials are characterised by their having effective permittivity or permeability tensors with principle components that differ in sign.
It is this feature that allows lenses created with hyperbolic metamaterials to break the diffraction limit.
More generally, in the setting of time-harmonic fields, hyperbolic metamaterials are governed by effective hyperbolic partial differential equations (PDEs) as opposed to elliptic PDEs for classical wave-like systems.

In the present paper we describe a new class of dynamic multi-scale structures, which we refer to as \emph{parabolic metamaterials}.
This novel group of materials is characterised by having effective partial differential equation of the parabolic type and can exhibit a spatially localised unidirectional wave propagation.
Similar effects have been observed in a different physical configuration, for flexural waves in periodically pinned Kirchhoff plates~\cite{mcphedran2015parabolic}, but were not associated with the notion of parabolic metamaterials.

The concept of parabolic metamaterials is associated with the presence of Dirac points on the dispersion surfaces.
Dirac points have been extensively studied in both the physics and mathematics literature (see~\cite{mei2012first,chan2012dirac}, among others), motivated by the novel electronic transport properties of graphene~\cite{neto2009electronic}.
Dirac cones have also recently been observed in phononic crystals by exploiting \emph{accidental degeneracies} caused by appropriately tuning material and geometric parameters in order to induce degeneracies at the centre of the Brillouin zone~\cite{chan2012dirac}.
In the present paper, the Dirac cones we exploit result from symmetries of the lattice system (so-called essential degeneracies).
These Dirac cones are connected by relatively flat narrow regions on the dispersion surfaces, which we term \emph{Dirac bridges}.
These Dirac bridges possess resonances where the dispersion surfaces are locally parabolic, which give rise to highly localised unidirectional wave propagation.

We emphasise that the effects described in the present paper are dynamic in nature and the parabolic metamaterial behaviour is associated with resonances of the system.
With this in mind, the novel method of high frequency homogenisation developed in a recent series of papers~\cite{craster2010high,craster2010higha,antonakakis2012high,antonakakis2014homogenisation,maling2015homogenisation,colquitt2015high}, will prove useful in the description and analysis of these parabolic metamaterials.
Classical multi-scale asymptotic homogenisation approaches for microstructured media (see~\cite{bensoussan2011asymptotic,panasenko2005multi} among others) usually involve the study of a class of static model problems on an elementary cell and can be considered valid in the long-wave quasi-static regime.
In contrast the approach of high frequency homogenisation considers perturbations away from known resonances and is therefore ideally suited to the study of dynamic parabolic wave propagation and the associated problems examined in the present paper.
The approach yields effective partial differential equations with constant coefficients that govern the dynamic macroscopic behaviour in the vicinity of resonances.
The methodology has proved very useful for a range of physical systems including continuous~\cite{antonakakis2014homogenisation} and discrete elasticity~\cite{colquitt2015high}, electromagnetism~\cite{maling2015homogenisation}, platonics~\cite{antonakakis2012high}, as well as scalar problems~\cite{craster2010high} where the approach has been particularly useful in the study of hyperbolic dispersion surfaces, dynamic anisotropy, and star-shaped waveforms~\cite{craster2012dangers}.

The concept of dynamic anisotropy and localisation in periodic lattices has been studied in several earlier papers~\cite{langley1996response,langley1997response,Ruzzene2003,ayzenberg2008resonant,osharovich2010wave,colquitt2012dynamic,movchan2014resonant}.
In the context of the theoretical studies~\cite{ayzenberg2008resonant,osharovich2010wave,colquitt2012dynamic,movchan2014resonant}, hyperbolic metamaterials can be viewed as multi-scale structures that support a class of Bloch-Floquet waves and exhibit locally hyperbolic behaviour in the neighbourhood of certain resonances.
In such cases, strong localisation can occur with wave propagation only permitted along directions associated with the principle curvatures of the hyperbolic surface.
The paper~\cite{colquitt2012dynamic} provides an important connection between the dispersive properties of Bloch-Floquet waves in an infinite lattice and the solutions of forced problems, which exhibit spatially localised waveforms of different types.

The present paper is developed as follows.
The formulation of the problem and discussion of dispersion surfaces are given in section \ref{S2}.
The notions of parabolic and hyperbolic metamaterials are discussed in section \ref{S3}, where we also discuss the method of high frequency homogenisation.
The theoretical discussion is complemented with numerical computations illustrating the highly localised wave propagation associated with parabolic metamaterials.
The paper is finalised in section \ref{S4} with some discussion and concluding remarks. 

\section{Formulation of the problem}
\label{S2}

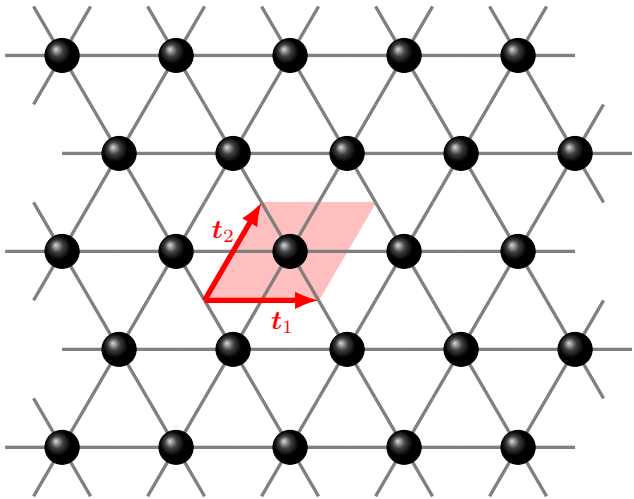
\begin{SCfigure}
\centering
\begin{tikzpicture}[scale=1.5]

\begin{scope}[shift={(-0.75,-0.25*3^(0.5))}]
	\draw[color = red!25, fill=red!25] (0,0) -- (1,0) -- (1.5,{0.5*3^(0.5)}) -- (0.5,{0.5*3^(0.5)}) -- (0,0);
\end{scope}

\foreach \i in {-2,-1,0,1,2} {
	\draw[line width=1.25, color=gray] (\i-0.5,0) -- (\i+0.5,0);
	\draw[line width=1.25, color=gray, rotate around={60:(\i,0)}] (\i-0.5,0) -- (\i+0.5,0);
	\draw[line width=1.25, color=gray, rotate around={120:(\i,0)}] (\i-0.5,0) -- (\i+0.5,0);
	\draw[shading=ball, ball color=black] (\i,0) circle (0.15);
}

\begin{scope}[shift={(0.5,0.5*3^(0.5))}]
\foreach \i in {-2,-1,0,1,2} {
	\draw[line width=1.25, color=gray] (\i-0.5,0) -- (\i+0.5,0);
	\draw[line width=1.25, color=gray, rotate around={60:(\i,0)}] (\i-0.5,0) -- (\i+0.5,0);
	\draw[line width=1.25, color=gray, rotate around={120:(\i,0)}] (\i-0.5,0) -- (\i+0.5,0);
	\draw[shading=ball, ball color=black] (\i,0) circle (0.15);
}

\end{scope}

\begin{scope}[shift={(0,3^(0.5))}]
\foreach \i in {-2,-1,0,1,2} {
	\draw[line width=1.25, color=gray] (\i-0.5,0) -- (\i+0.5,0);
	\draw[line width=1.25, color=gray, rotate around={60:(\i,0)}] (\i-0.5,0) -- (\i+0.5,0);
	\draw[line width=1.25, color=gray, rotate around={120:(\i,0)}] (\i-0.5,0) -- (\i+0.5,0);
	\draw[shading=ball, ball color=black] (\i,0) circle (0.15);
}

\end{scope}

\begin{scope}[shift={(0.5,-0.5*3^(0.5))}]
\foreach \i in {-2,-1,0,1,2} {
	\draw[line width=1.25, color=gray] (\i-0.5,0) -- (\i+0.5,0);
	\draw[line width=1.25, color=gray, rotate around={60:(\i,0)}] (\i-0.5,0) -- (\i+0.5,0);
	\draw[line width=1.25, color=gray, rotate around={120:(\i,0)}] (\i-0.5,0) -- (\i+0.5,0);
	\draw[shading=ball, ball color=black] (\i,0) circle (0.15);
}

\end{scope}

\begin{scope}[shift={(0,-3^(0.5))}]
\foreach \i in {-2,-1,0,1,2} {
	\draw[line width=1.25, color=gray] (\i-0.5,0) -- (\i+0.5,0);
	\draw[line width=1.25, color=gray, rotate around={60:(\i,0)}] (\i-0.5,0) -- (\i+0.5,0);
	\draw[line width=1.25, color=gray, rotate around={120:(\i,0)}] (\i-0.5,0) -- (\i+0.5,0);
	\draw[shading=ball, ball color=black] (\i,0) circle (0.15);
}
\end{scope}

\begin{scope}[shift={(-0.75,-0.25*3^(0.5))}]
	\draw[line width = 2, color = red, -latex] (0,0) -- (1,0);
	\draw[line width = 2, color = red, -latex] (0,0) -- (0.5,{0.5*3^(0.5)});
	\node[right] at (0.5,-0.2) {$\displaystyle{\color{red}\vec{t}_1}$};
	\node[left] at (0.35,{0.35*3^(0.5)}) {$\displaystyle{\color{red}\vec{t}_2}$};
\end{scope}

\end{tikzpicture}
\caption{\label{fig:lattice-schematic}
The uniform triangular lattice with the black balls denoting concentrated masses and the grey lines massless elastic beams.
The primitive cell is shaded in red and the direct lattice vectors are shown, also in red.}
\end{SCfigure}

In the present paper we study the in-plane motion of an elastic triangular lattice which, initially, might seem to be a conventional and well studied mechanical system.
Conventionally such structures would be considered to be so-called \emph{``stretch dominated''}~\cite{deshpande2001foam}, that is, the lattice remains statically determinate if the flexural rigidity of the links is neglected.
Therefore, the bending effects in asymptotic models of thin  elastic ligaments are often neglected for triangular lattices, both in statics and dynamics.
However, here we show that this asymptotic approximation requires an adjustment in the dynamic regime and the effects of bending, generally neglected, play a fundamental role leading to a novel class of metamaterial behaviour.
In the present paper we introduce the notion of a new type of elastic metamaterials referred to as {\em parabolic metamaterials}.

The problem is formulated using the same framework as established in~\cite{colquitt2012dynamic,colquitt2015high}.
The geometry is shown in figure~\ref{fig:lattice-schematic}, the elastic ligaments (denoted by grey links) are treated as massless Euler-Bernoulli beams, which allow for both flexural and extensional modes, and the junction points have both translational and rotational inertia.
It is convenient, without loss of generality, to normalise the system such that both the length of the ligaments and the mass of each junction is unity.
Each node is labelled by the integer multi-index $\vec{m}\in\mathbb{Z}^2$ such that the spatial position of the $\vec{m}^{\text{th}}$ node is $\vec{x}(\vec{m}) = \mathsf{T}\vec{m}$, where $\mathsf{T} = [\vec{t}_1,\vec{t}_2]$ is matrix with columns formed from the direct lattice vectors $\vec{t}_1 = [1,0]^\mathrm{T}$ and $\vec{t}_2 = [1/2,\sqrt{3}/2]^\mathrm{T}$.
Each node is connected to its set of nearest neighbours, denoted by $\mathcal{N} = \{\vec{p}:\|\vec{x}(\vec{m}+\vec{p}) - \vec{x}(\vec{m})\| \leq 1 \}$.
We emphasise that $\vec{0}\in\mathcal{N}$, i.e. each node is connected to itself.
The inertial properties of each node are described by the matrix $\mathsf{M} = \text{diag}[1,1,J]$ where $J$ is the polar moment of inertia of each node.
Finally, the elastic stiffness of the ligament connecting node $\vec{m}+\vec{p}$ to node $\vec{m}$ is described by the matrix $\mathsf{C}(\vec{p})$.
Using this notation, the equations of motion for time-harmonic waves propagating in the lattice can be conveniently written in the form
\begin{equation}
\omega^2\mathsf{M}\vec{u}(\vec{m}) = \sum_{\vec{p}\in\mathcal{N}} \mathsf{C}(\vec{p}) \vec{u}(\vec{m}+\vec{p}),
\label{eq:eom}
\end{equation}
where $\vec{u}(\vec{m}) = [u_1(\vec{m}),u_2(\vec{m}),\theta(\vec{m})]$ is the generalised displacement of each node: $u_1$ and $u_2$ are the usual Cartesian displacements and $\theta$ is the angle of in-plane rotation.
The stiffness matrices $\mathsf{C}(\vec{p})$ can be deduced from those given in our earlier paper~\cite[Eq. 9]{colquitt2012dynamic} by taking the limit $\varrho\to0$.
For the convenience of the reader, we also provide them in appendix~\ref{app:matrices}.
We introduce the discrete Fourier transform and its inverse
\[
\vec{u}^\mathrm{F}(\vec{k}) = \sum_{\vec{m}\in\mathbb{Z}^2} \vec{u}(\vec{m})e^{i\vec{k}\cdot\mathsf{T}\vec{m}},\quad
\vec{u}(\vec{m}) = \frac{1}{\|\mathcal{R}\|}\iint\limits_\mathcal{R} \vec{u}^\mathrm{F}(\vec{k})e^{-i\vec{k}\cdot\mathsf{T}\vec{m}}\,\mathrm{d}\vec{k},
\]
where $\mathcal{R} = (-\pi,\pi)\times(-2\pi/\sqrt{3},2\pi/\sqrt{3})$ for the uniform triangular lattice under consideration.
Applying the discrete Fourier transform to the equations of motion~\eqref{eq:eom}, we find
\begin{equation}
\left[\omega^2\mathsf{M} - \sum_{\vec{p}\in\mathcal{N}} \mathsf{C}(\vec{p})e^{i\vec{k}\cdot\mathsf{T}\vec{p}}\right]\vec{u}^\mathrm{F}(\vec{k}) = 0,
\label{eq:eom-ft}
\end{equation}
whence the dispersion equation is immediately obtained
\begin{equation}
\det\left[\omega^2\mathsf{M} - \sum_{\vec{p}\in\mathcal{N}} \mathsf{C}(\vec{p})e^{i\vec{k}\cdot\mathsf{T}\vec{p}}\right] = 0.
\label{eq:dispersion}
\end{equation}
The dispersion equation is a cubic polynomial in $\omega^2$ and its roots can therefore be expressed in closed form; the expressions for the roots are somewhat tedious and are therefore omitted for brevity.
The dispersion surfaces shown in figure~\ref{fig:dispersion} are a visual representation of the roots of the dispersion equation and show the frequency $\omega$ as a function of Bloch vector $\vec{k}$.

\subsection{Dispersion surfaces}

\begin{figure}
\centering
\begin{subfigure}[c]{0.45\linewidth}
\includegraphics[width=\linewidth]{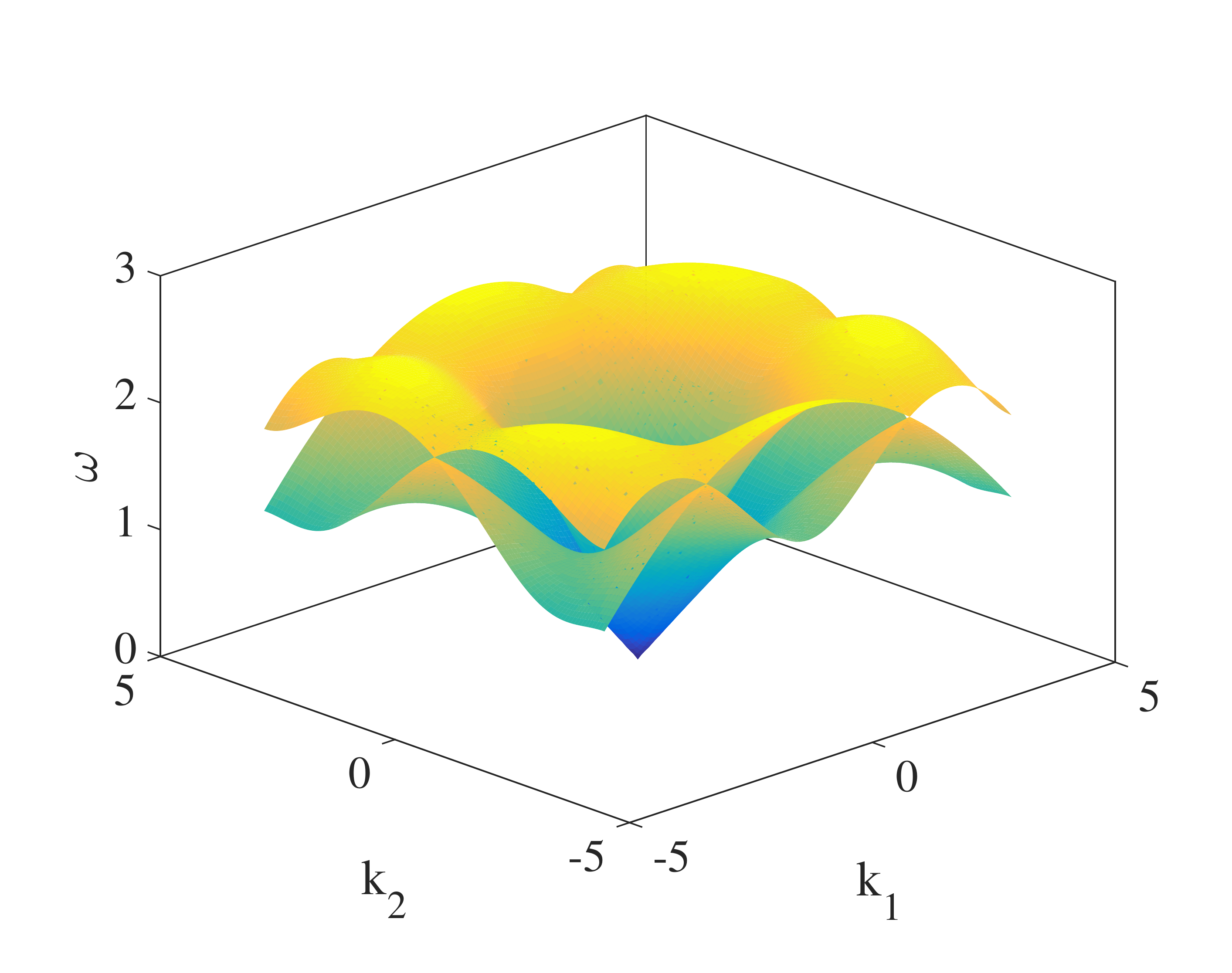}
\caption{\label{fig:truss_dispersion}}
\end{subfigure}
\begin{subfigure}[c]{0.45\linewidth}
\includegraphics[width=\linewidth]{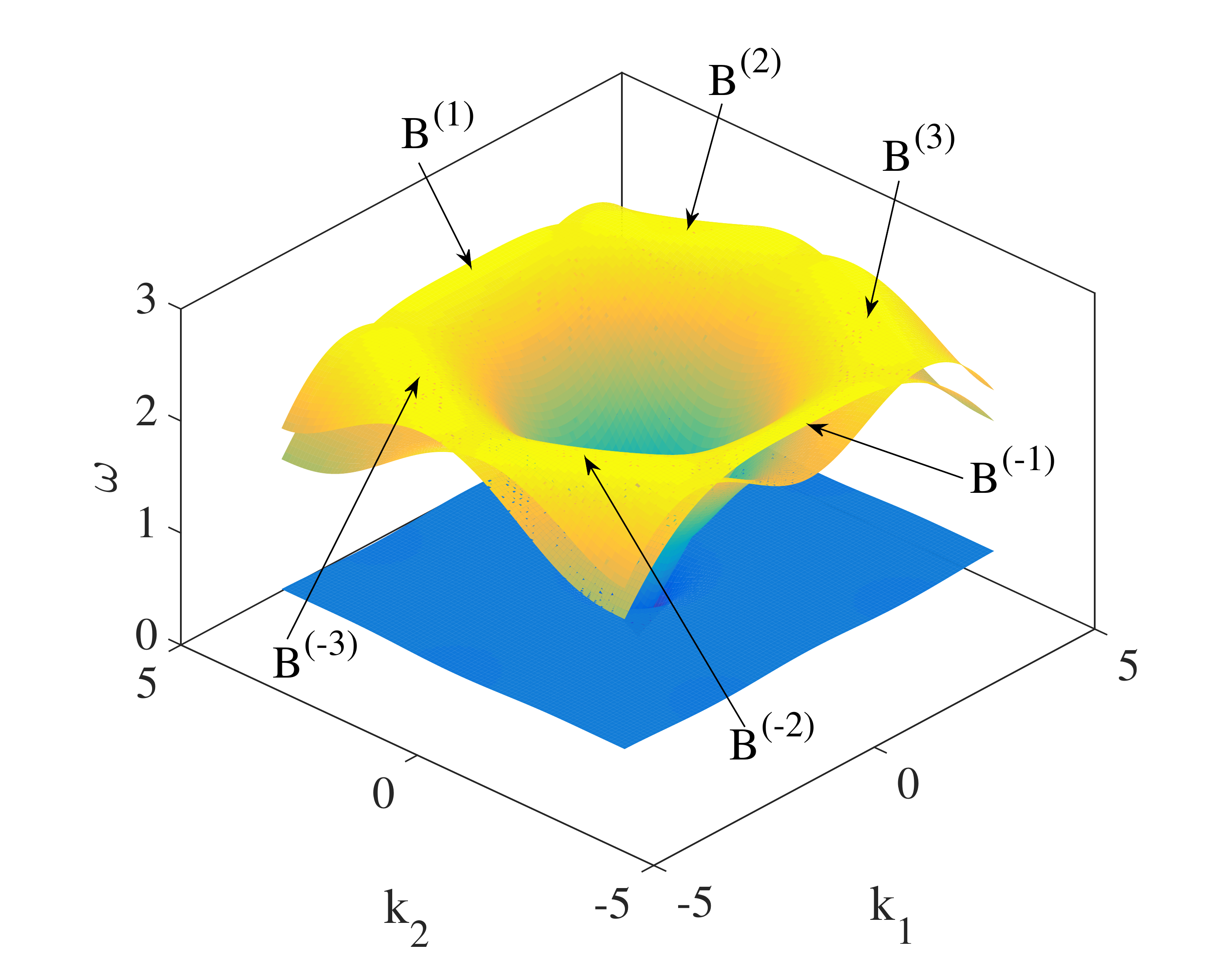}
\caption{\label{fig:flexural_dispersion}}
\end{subfigure}
\begin{subfigure}[c]{0.45\linewidth}
\vspace{0.025\linewidth}
\includegraphics[width=\linewidth]{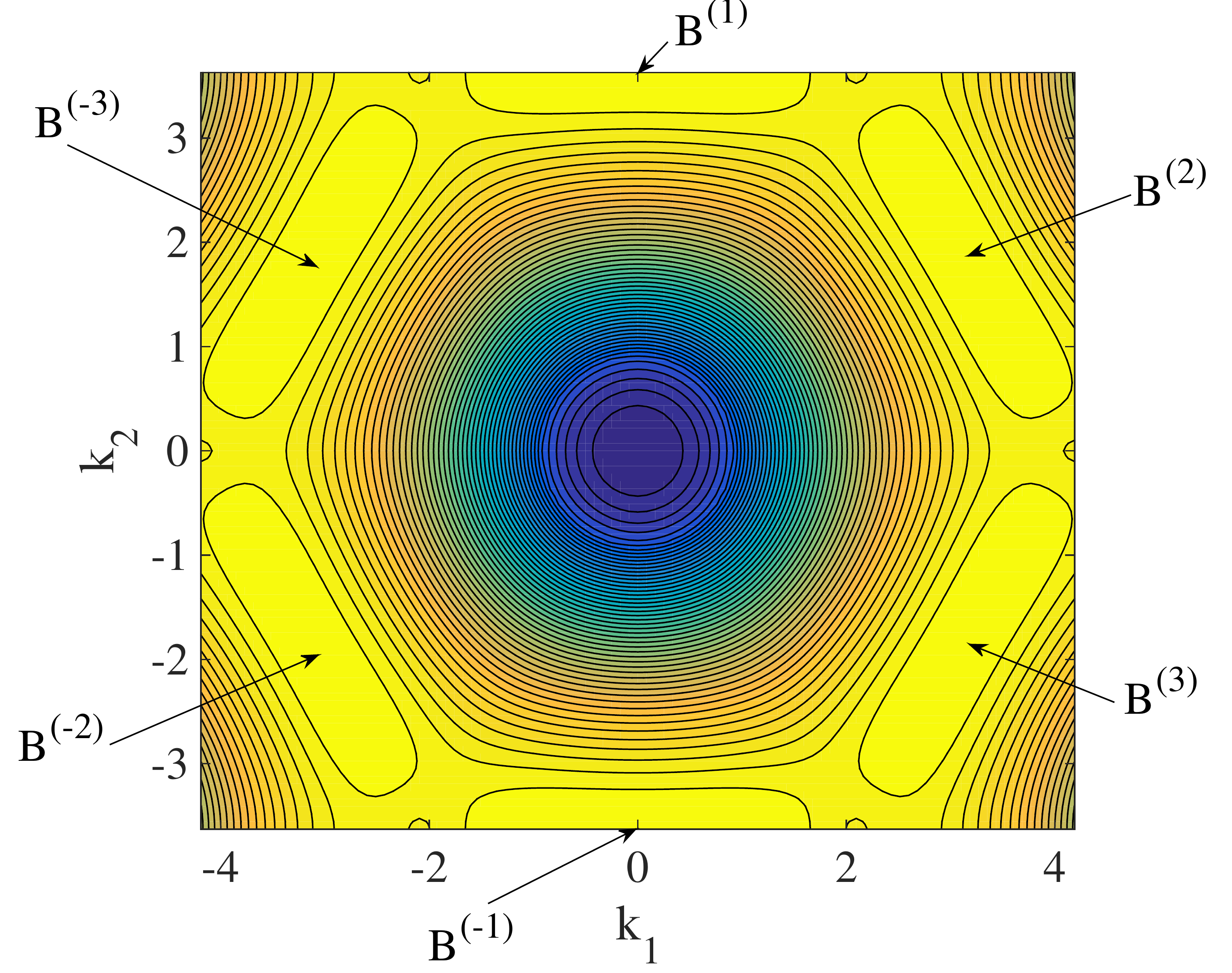}
\vspace{0.025\linewidth}
\caption{\label{fig:flexural_contour}}
\end{subfigure}
\begin{subfigure}[c]{0.45\linewidth}
\vspace{0.025\linewidth}
\includegraphics[width=\linewidth]{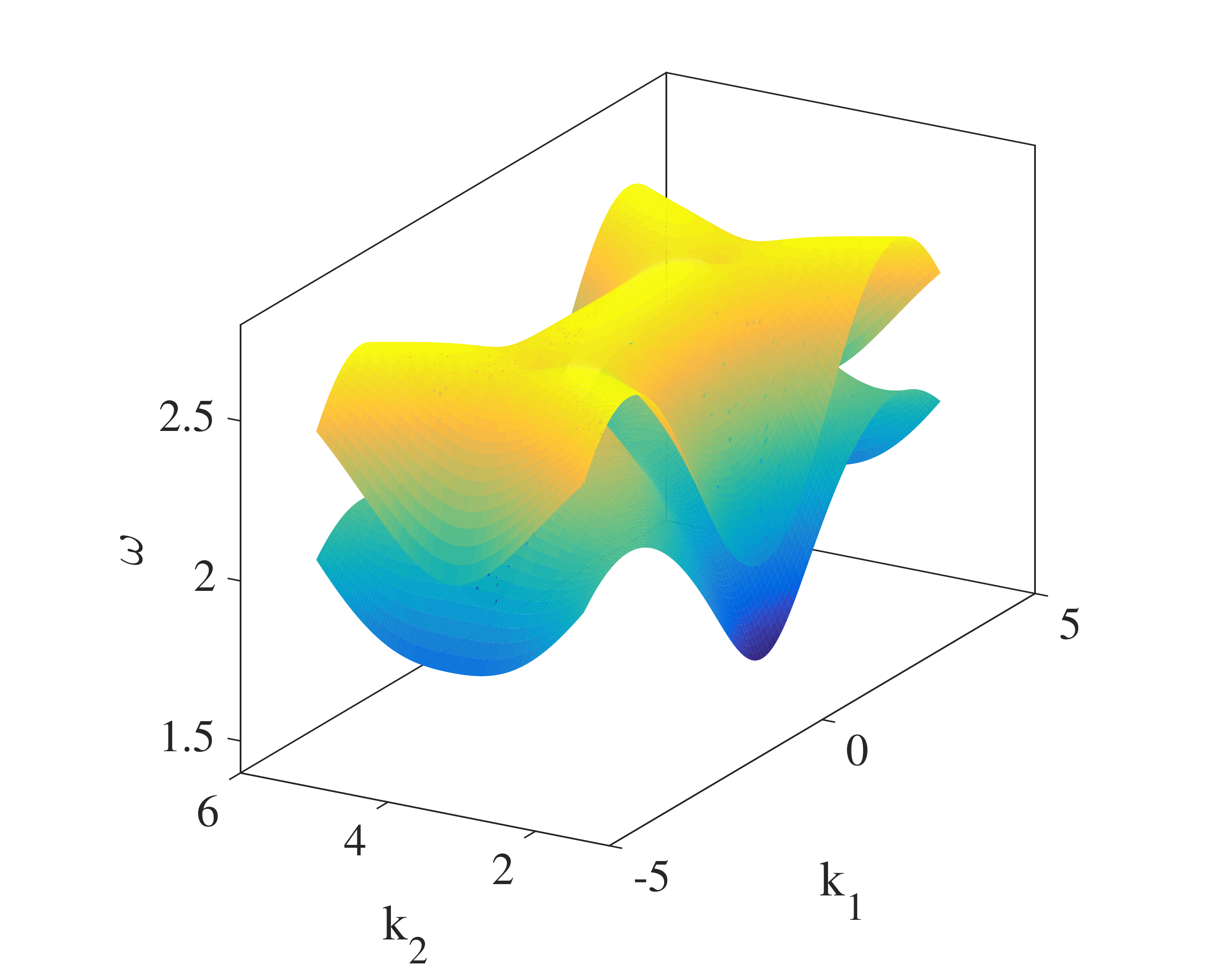}
\vspace{0.025\linewidth}
\caption{\label{fig:bridge_zoom}}
\end{subfigure}
\caption{\label{fig:dispersion}
The dispersion surfaces (a), (b), and (d) and slowness contours (c) for the triangular lattice.
Part (a) shows the dispersion curves for the triangular lattice where only the compressional modes of the elastic ligaments are accounted for.
These should be contrasted with part (b), which shows the surfaces for the case when both the flexural and compressional modes of the elastic ligaments are accounted for.
The resonant points $B^{(\pm j)}$ at the centre of the Dirac bridges labelled.
The Dirac bridges are also clearly visible on the slowness contours shown in part (c).
A enlarged section of the highest two surfaces showing the Dirac bridge containing the locally parabolic point $B^{(1)}$ is shown in part (d).
The numerical values are $J = 2$ and $\beta = 0.071$.}
\end{figure}

Figure~\ref{fig:dispersion} shows two cases: the first, in panel~\ref{fig:truss_dispersion}, is for the case when the flexural modes of the elastic ligaments are neglected and only extensional modes are considered; the second, in panel~\ref{fig:flexural_dispersion}, shows the same configuration but with the bending modes of the links taken into account.
The additional, predominantly flat band, shown in~\ref{fig:flexural_dispersion} is a consequence of the bending modes of the ligaments and represents a rotationally dominant mode.
It should be emphasised that, in the former case where only extensional modes are considered, the shape of the dispersion surfaces is essentially determined by the geometry.
Altering the stiffness of the elastic ligaments and inertia of the junctions would fundamentally only act to stretch the dispersion surfaces.
This is in contrast to the latter case where the flexural deformation of the elastic ligaments is taken into account.
In this non-dimensional form there are two free parameters which affect the shape and curvature of the dispersion surfaces: $\beta$, roughly speaking the ratio of bending stiffness to extensional stiffness of the ligaments, and $J$, which characterises the ratio of rotational inertia to translational inertia of the junctions.
In terms of physical parameters, $\beta = 2I/(S\ell^2)$ and $J = \bar{J}/(m\ell^2)$ where $m$ and $\bar{J}$ are the mass and polar moment of inertia of the lattice junctions respectively; $I$, $S$, and $\ell$ are the area moment of inertia, cross-sectional area, and length of the lattice ligaments respectively.

These surfaces have several interesting features including the characteristic semi-infinite stop band associated with discrete structures, flat bands which can give rise to slow waves, and the presence of Dirac cones that have now become synonymous with hexagonal and triangular geometries.
However our interest lies with a novel feature of the dispersion surfaces that, although related to Dirac cones, has received little attention.
These features are only present in the case when the flexural effects of the elastic ligaments are included, as shown in figure~\ref{fig:flexural_dispersion}.
In particular, it will be shown that the points indicated by $B^{(\pm j)}$ correspond to resonances and, at these points, the dispersion surfaces are locally parabolic leading to what we term \emph{parabolic metamaterial} effects.
These parabolic regions connect neighbouring Dirac points, hence we refer to them as \emph{Dirac bridges}.

\section{Constructing Dirac bridges}
\label{S3}

The presence of Dirac cones is generally associated with symmetries of the system through its geometry (see, for example,~\cite{mei2012first}) and therefore we would always expect regular triangular geometries to possess Dirac cones.
The presence of parabolic Dirac bridges, in contrast, requires both the existence of Dirac cones in addition to the careful tuning of material parameters.
Although one can obtain explicit expressions for the dispersion surfaces and, hence, their curvature in closed form, these expressions are often tedious and it can be difficult to extract any physical insight from them.
Here, we employ the recently developed theory of high frequency homogenisation~\cite{craster2010high,craster2010higha}.
The multiple-scales approach, originally developed in~\cite{craster2010high}, involves perturbations away from resonances and provides the effective continuous partial differential equation that governs the long scale behaviour of the medium.
The rapid oscillations associated with the resonance and multiple scattering are encoded in the material parameters of the effective medium and reflected in the coefficients of the homogenised PDE.
The approach has recently been extended to elastic lattices~\cite{colquitt2015high}, which makes it an ideal tool to analyse Dirac bridges for the structures considered here.

\subsection{High frequency homogenisation}
\label{sec:hfh}

We will now briefly summarise the essential features of the high frequency homogenisation methodology.
For further details and a complete description of the approach, the reader is referred to~\cite{colquitt2015high} and reference therein.

We begin by introducing two scales: the short-scale discrete variable $\vec{m}$, and the long-scale continuous variable $\vec{\eta} = \epsilon\mathsf{T}\vec{m}$.
It is assumed that the small parameter $0 < \epsilon \ll 1$ characterises the short-scale of the lattice.
For example, if the lattice is formed by an $N\times N$ grid of particles, where $N\gg1$, then $\epsilon = 1/N$.
The displacement is then considered as a function of two independent vector-valued variables: $\vec{u} = \vec{u}(\vec{m},\vec{\eta})$.
Expanding the Fourier-transformed equations of motion~\eqref{eq:eom-ft} in the continuous variable $\vec{\eta}$ for small $\epsilon$ we find
\[
\omega^2\mathsf{M}\vec{u}^\mathrm{F}(\vec{k},\vec{\eta}) = \left[
\sigma_0(\vec{k}) + \epsilon\sigma_1(\vec{k}) + \epsilon^2\sigma_2(\vec{k}) + \mathcal{O}(\epsilon^3)\right]\vec{u}^\mathrm{F}(\vec{k},\vec{\eta}),
\]
where
\begin{align}
\sigma_0(\vec{k}) & = \sum_{\vec{p}\in\mathcal{N}} \mathsf{C}(\vec{p})e^{i\vec{k}\cdot\mathsf{T}\vec{p}}, \\
\sigma_1(\vec{k},\nabla) & = \sum_{\vec{p}\in\mathcal{N}} \mathsf{C}(\vec{p})e^{i\vec{k}\cdot\mathsf{T}\vec{p}}\mathsf{T}\vec{p}\cdot\nabla, \\
\sigma_2(\vec{k},\nabla) & = \frac{1}{2}\sum_{\vec{p}\in\mathcal{N}} \mathsf{C}(\vec{p})e^{i\vec{k}\cdot\mathsf{T}\vec{p}}\mathsf{T}\vec{p}\cdot\left[\mathsf{T}\vec{p}\cdot\nabla\nabla\right],
\end{align}
and $\nabla$ acts on the continuous long-scale variable $\vec{\eta}$.
Next, we take the following Ans\"{a}tze for the Fourier-transformed displacement and frequency-squared
\[
\vec{u}^\mathrm{F}(\vec{k},\vec{\eta}) = \sum_{n=0}^\infty \epsilon^n\vec{u}^\mathrm{F}_n(\vec{k},\vec{\eta}),\quad
\omega^2 = \sum_{n=0}^\infty \epsilon^n\omega^2_n,
\]
which yields a hierarchy of algebraic problems in ascending orders of $\epsilon$.
For most of what follows, it is necessary to consider only the first three problems,
\begin{align}
\label{eq:zero-order}
\mathsf{S}_0(\omega,\vec{k})\vec{u}^\mathrm{F}_0(\vec{k},\vec{\eta}) & = \vec{0},\\
\label{eq:first-order}
\mathsf{S}_0(\omega,\vec{k})\vec{u}^\mathrm{F}_1(\vec{k},\vec{\eta}) & = -\mathsf{S}_1(\omega,\vec{k},\nabla)\vec{u}^\mathrm{F}_0(\vec{k},\vec{\eta}),\\
\label{eq:second-order}
\mathsf{S}_0(\omega,\vec{k})\vec{u}^\mathrm{F}_2(\vec{k},\vec{\eta}) & = -\mathsf{S}_2(\omega,\vec{k},\nabla)\vec{u}^\mathrm{F}_0(\vec{k},\vec{\eta}) -\mathsf{S}_1(\omega,\vec{k},\nabla)\vec{u}^\mathrm{F}_1(\vec{k},\vec{\eta}),
\end{align}
where $\mathsf{S}_i = \omega_i^2\mathsf{M} - \sigma_i$.
Considering the leading order problem~\eqref{eq:zero-order}, we observe that $\mathsf{S}_0$ is independent of the long-scale variable $\vec{\eta}$ and its derivatives.
Therefore fixing $\vec{k}$, which is equivalent to setting the phase-shift across the elementary cell, reduces~\eqref{eq:zero-order} to an eigenvalue problem which, for simple eigenvalues, admits solutions of the form
\begin{equation}
\vec{u}^\mathrm{F}_0(\vec{k},\vec{\eta}) = \vec{U}_0^\mathrm{F}(\vec{k})\phi_0(\vec{\eta}),
\end{equation}
where we normalise such that $|\vec{U}_0^\mathrm{F}| = 1$.
Indeed, for the present configuration, it will be sufficient to only consider non-degenerate eigenvalues.
Examining the first order problem~\eqref{eq:first-order}, we find that it is solvable if and only if (see~\cite{colquitt2015high})
\[
{\vec{U}_0^\mathrm{F}(\vec{k})}^\dagger\mathsf{S}_1(\omega,\vec{k},\nabla)\vec{u}^\mathrm{F}_0(\vec{k},\vec{\eta}) = 0,
\]
where the dagger denotes the Hermitian Transpose.
For the cases of non-degenerate eigenvalues considered in the present paper we find that ${\vec{U}_0^\mathrm{F}(\vec{k})}^\dagger\sigma_1(\vec{\omega},\vec{k},\nabla)\vec{U}_0^\mathrm{F}(\vec{k}) = 0$ which implies $\omega_1^2 = 0$ and the first order problem admits the solution
\[
\vec{u}_1^\mathsf{F}(\vec{k},\vec{\eta}) = \mathsf{S}_0^+\sigma_1(\omega,\vec{k},\nabla)\vec{u}_0^\mathsf{F}(\vec{k},\vec{\eta}) + \left[
\mathsf{I} - \mathsf{S}_0^+\mathsf{S}_0\right]\vec{\psi}(\vec{k},\vec{\eta}),
\]
where the superscript plus sign denotes the Moore-Penrose pseudoinverse, $\vec{\psi}$ is an arbitrary vector, and $\mathsf{I}$ is the identity matrix of appropriate dimension.

Moving to the second order problem, applying the solvability condition, and using the fact that $\left[\mathsf{I} - \mathsf{S}_0^+\mathsf{S}_0\right]$ is the orthogonal projector onto the kernel of $\mathsf{S}_0$, we obtain the second order partial differential equation for the leading order envelope function $\phi_0(\vec{\eta})$ and the correction to the frequency
\begin{equation}
\left[{\vec{U}_0^\mathrm{F}}^\dagger\sigma_1\mathsf{S}_0^+\sigma_1\vec{U}_0^\mathrm{F} +
{\vec{U}_0^\mathrm{F}}^\dagger\mathsf{S}_2{\vec{U}_0^\mathrm{F}} - \omega_2^2\right]\phi_0(\vec{\eta}) = 0,
\label{eq:homogenised-pde}
\end{equation}
where we suppress the arguments of the operators.
Equation~\eqref{eq:homogenised-pde} represents the homogenised partial differential equation that governs the long-scale response of the medium.
We emphasise that the envelope function in~\eqref{eq:homogenised-pde} is a function of the long-scale continuous variable $\vec{\eta}$ only.
The short-scale behaviour is encapsulated by the bracketed operator in~\eqref{eq:homogenised-pde}; this operator corresponds to a second order partial differential equation with constant coefficients and will be used in the following section to construct the parabolic metamaterial.

\subsection{Hyperbolic, elliptic, and parabolic metamaterials}

We now return to the points $B^{(\pm j)}$, which lie on the boundary of the Brillouin zone.
By considering the symmetries of the lattice, we can restrict our attention to the irreducible Brillouin zone formed by a triangle with vertices at $\vec{k} = [0,0]^\mathrm{T}$, $[0,2\pi/\sqrt{3}]^\mathrm{T}$, and $[2\pi/3,2\pi/\sqrt{3}]^\mathrm{T}$.
Only $B^{(1)}$ lies within the closure of the first Brillouin zone; the remaining points can be obtained by symmetry operations.
Therefore, the following will be restricted to analysis of the point $B^{(1)}$.
We are primarily interested in the behaviour of the upper two surfaces at this point as these exhibit the desired locally parabolic behaviour.
In particular, on these surfaces and for certain values of $\beta$ and $J$, the point $B^{(1)}$ lies at the centre of the Dirac bridge which connects adjacent Dirac points.

Using the approach outlined in \S\ref{sec:hfh}, we now proceed to derive the effective partial differential equations that govern the response of the lattice in the neighbourhood of $B^{(1)}$.
The leading order problem~\eqref{eq:zero-order} has three eigenvalues
\[
\Omega_{(0,1)}^2 = \frac{10\beta}{J},\quad
\Omega_{(0,2)}^2 = 2 + 36\beta,\quad\text{and}\quad
\Omega_{(0,3)}^2 = 6+12\beta,
\]
with the corresponding short-scale eigenvectors $\vec{U}_{(0,1)} = [0,0,1]^\mathrm{T}$, $\vec{U}_{(0,2)} = [1,0,0]^\mathrm{T}$, and $\vec{U}_{(0,3)} = [0,1,0]^\mathrm{T}$.
For thin elastic ligaments of length $\ell$ and thickness $d$, $\beta = \mathcal{O}(d^2/\ell^2)$ with $d/\ell \ll1$; and, typically, $J=\mathcal{O}(1)$.
Therefore, the first eigenvalue corresponds to the lower dispersion surface with the latter two associated with the upper two surfaces.
Proceeding to higher order, we find that the second and third modes yield effective partial differential equations of the form
\begin{equation}
\left[a^{(p)}_{ij}\frac{\partial^2}{\partial \eta_i\partial \eta_j} - \omega^2_{(2,p)}\right] \phi_0(\vec{\eta}) = 0,
\label{eq:homo-pde}
\end{equation}
where $p = 1,2,3$ and enumerates the modes.
The bracketed differential operator in~\eqref{eq:homo-pde} is the explicit form of the same bracketed operator appearing in~\eqref{eq:homogenised-pde}.
The matrices $A^{(p)}$ are all diagonal and have the following forms
\begin{equation}
\label{eq:A1}
A^{(1)} = \frac{\beta}{2}
\begin{bmatrix}
[5(3J - 1)\beta + 3J]/\alpha_1 & 0 \\
0 & 3[(9J + 5)\beta - J]/\alpha_2
\end{bmatrix},
\end{equation}
\begin{equation}
\label{eq:A2}
A^{(2)} = \frac{1}{8}
\begin{bmatrix}
18\beta - 7 & 0 \\
0 & 3[(324J - 198)\beta^2 + (36J - 5)\beta + J]/\alpha_2
\end{bmatrix}
\end{equation}
\begin{equation}
\label{eq:A3}
A^{(3)} = \frac{3}{8}
\begin{bmatrix}
[(58 - 84J)\beta^2 + (5 - 36J)\beta - 3J]/\alpha_1 & 0 \\
0 & 3(1 + 2\beta),
\end{bmatrix},
\end{equation}
where $\alpha_1 = (18J - 5)\beta + J$ and $\alpha_2 = (6J - 5)\beta + 3J$.
Upon examination of~\eqref{eq:A1}--\eqref{eq:A3}, it can be shown that the eigenvalues of $A^{(p)}$ can have different signs depending on the values of $\beta$ and $J$, leading to effective partial differential equations.
As remarked above, the first mode is typically associated with the lowest frequency band and although the desired locally parabolic behaviour is present, it will occur at frequencies where several other modes exist making the parabolic behaviour difficult to isolate.
We therefore focus on the higher two modes for $p=2,3$.

In order the classify the different behaviours expected from the homogenised partial differential equations, we introduce the following definition, that is consistent with the classical classification of linear second order partial differential equations.
\begin{definition}
Defining the discriminant of the homogenised partial differential equation~\eqref{eq:homo-pde} as $\Delta^{(p)} = \det A^{(p)}$, the partial differential equation is then said to be
\begin{enumerate}
\item\textbf{Elliptic} if $\Delta < 0$,
\item\textbf{Parabolic} if $\Delta = 0$,
\item\textbf{Hyperbolic} if $\Delta > 0$.
\end{enumerate}
\end{definition}
Figure~\ref{fig:parameter_space} shows a plot of $\sgn(\Delta^{(p)})$ for a range of $(J,\beta)$ values, illustrating the different regimes.
The grey (white) regions indicate that the effective partial differential equation is hyperbolic (elliptic).
These regions are demarcated by solid black curves where the effective partial differential equation is parabolic.
By tuning the material parameters of the triangular lattice, we can control the effective dynamic behaviour of the medium.
For monochromatic waves, elliptic equations represent `normal' wave propagation (e.g. the Helmholtz equation for acoustic waves), hyperbolic equations result in highly localised waves along a discrete set of principle directions (see, for example,~\cite{colquitt2012dynamic}), and parabolic equations are associated with strongly localised beams where waves are only permitted to propagate along a single principle direction.

\subsubsection{Parabolic profiles}

\begin{figure}
\centering
\begin{subfigure}[c]{0.45\linewidth}
\includegraphics[width=\linewidth]{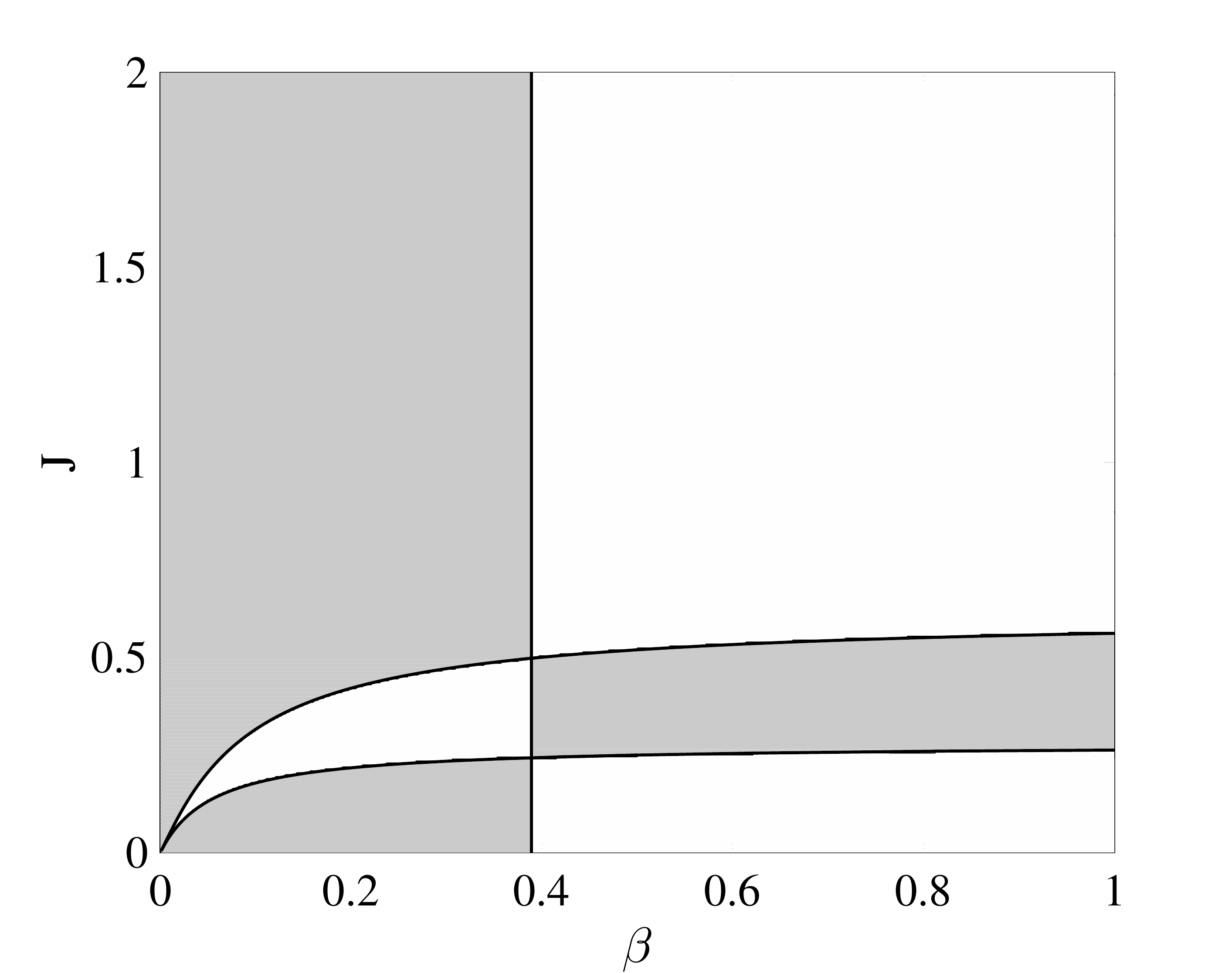}
\caption{\label{fig:parameter_space_mode2}}
\end{subfigure}
\begin{subfigure}[c]{0.45\linewidth}
\includegraphics[width=\linewidth]{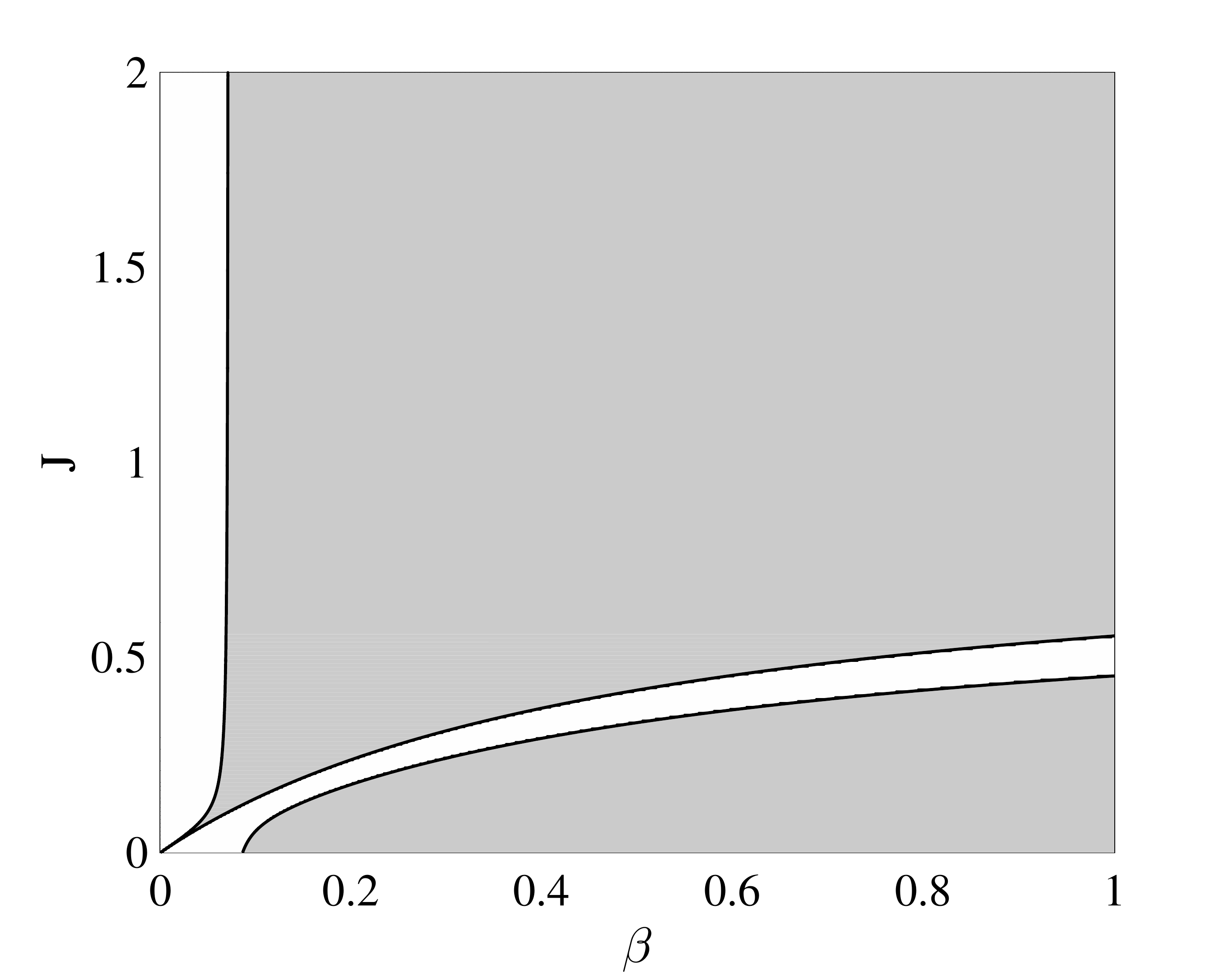}
\caption{\label{fig:parameter_space_mode3}}
\end{subfigure}
\caption{\label{fig:parameter_space}
Here we plot the sign of the discriminant of the effective partial differential equation for the triangular lattice over a range of $(\beta, J)$ values.
The white regions correspond to parameter values which yield elliptic behaviour; whereas grey regions correspond to effective PDEs of hyperbolic type.
These regions are separated by solid lines where the discriminant of the effective PDEs vanish and the lattice behaves as a parabolic metamaterial.
Part (a) shows the sign of the discriminant for the second mode, whilst part (b) corresponds to the third mode.}
\end{figure}

Parabolic partial differential equations are characterised by vanishing discriminants, which implies that one of the eigenvalues of $A^{(p)}$ is zero.
These eigenvalues are quadratic polynomials in $\beta$ and therefore we can obtain their roots explicitly.
Since $\beta$ and $J$ represent stiffness and mass respectively, we only consider positive roots.
In particular, for the second eigenmode the effective PDE is parabolic when
\[
\beta^{(2)}_{11} = \frac{7}{18}\quad\text{and}\quad
\beta^{(2)}_{22} = \frac{36J - 5 + \sqrt{432J + 25}}{36(11 - 18J)},
\]
where $\beta^{(2)}_{jj}$ indicates that the coefficient of $\partial^2/\partial\eta_j^2$ vanishes for this value of $\beta$.
Both $\beta$ and $J$ correspond to physical quantities (roughly speaking the ratio of flexural to compressional stiffness of the elastic ligaments and the ratio of translational to rotational inertia respectively) and therefore should take positive finite values.
The branch of the square root is chosen accordingly.
For the third eigenmode we have similar results
\[
\beta^{(3)}_{(11,1)} = \frac{\sqrt{25 - 336J + 2304J^2} - 5 - 36J}{4(42 J - 29)},\quad
\beta^{(3)}_{(11,2)} = \frac{\sqrt{25 - 336J + 2304J^2} + 5 + 36J}{4(29 - 42 J)},
\]
In order to satisfy the condition that $\beta > 0$, the second root $\beta^{(3)}_{(11,2)}$ is restricted to $0 < J < 29/42$.
These values are indicated by the solid black curves on figure~\ref{fig:parameter_space} and demark the boundary between the regions where the effective PDE is either hyperbolic or elliptic.
It is interesting to note that $\beta^{(2)}_{11}$ is independent of $J$ which suggests that, for this class of lattice, one can obtain parabolic behaviour by simply tuning the stiffness of the elastic ligaments without altering the inertial properties thus simplifying design and implementation of parabolic metamaterials.

Special attention is required at the poles of the tensors $A^{(i)}$, that is, when $\alpha_1 = 0$ and/or $\alpha_2 = 0$.
When $\alpha_1 = 0$, the first and third eigenmodes coincide whilst the first and second eigenmodes coincide when $\alpha_2 = 0$.
All three modes coincide when $\alpha_1 = \alpha_2 = 0$ ($J = 24/5$).
In such cases of degenerate eigenvalues, the leading order solution is a linear combination of the eigensolutions for each eigenvalue
\[
\vec{u}_0^\mathsf{F}(\vec{k},\vec{\eta}) = \sum_{i \in \mathcal{D}} \vec{U}_{(0,i)}^\mathrm{F}(\vec{k})\phi_{(0,i)}(\vec{\eta}),
\]
where $\vec{U}_{(0,i)}^\mathrm{F}(\vec{k})$ is the eigenvector corresponding to $\omega_{(0,i)}^2$, $\phi_{(0,i)}(\vec{\eta})$ is the associated longscale envelope function, and $\mathcal{D}$ is the set of indices enumerating the degenerate modes.
The order of degeneracy is then $\|\mathcal{D}\|$.
The solvability conditions for the first order problem are then
\[
{\vec{U}_{(0,i)}^\mathrm{F}(\vec{k})}^\dagger\mathsf{S}_1(\omega,\vec{k},\nabla)\vec{u}^\mathrm{F}_{(0,j)}(\vec{k},\vec{\eta}) = 0,\quad\text{for}\;i,j \in \mathcal{D},\]
which together yield a coupled system of $\|\mathcal{D}\|$ first order partial differential equations for the longscale envelope functions and the first order correction to the frequency.
In some cases the system may be explicitly decoupled.
For example, when $\alpha_1 = 0$ and the first and third modes coincide, we find that the coupled system reduces to a single uncoupled parabolic partial differential equation governing the envelope functions for the first and third eigenmodes
\begin{equation}
\left( \frac{9}{6J - 5} \right)^2 \frac{\partial^2}{\partial\eta_2^2}\phi_{(0,i)} + \omega_1^4\phi_{(0,i)} = 0, \quad\text{for}\;i=1,3.
\end{equation}
This case ($\alpha_1 = 0$) is associated with the lower solid curve in figure~\ref{fig:parameter_space_mode2}.
Similarly, when $\alpha_2 = 0$ the first and second modes coincide and we find that the associated envelope functions are governed by another parabolic equation
\begin{equation}
\left( \frac{9}{18J - 5} \right)^2 \frac{\partial^2}{\partial\eta_2^2}\phi_{(0,i)} + \omega_1^4\phi_{(0,i)} = 0, \quad\text{for}\;i=1,2.
\end{equation}
This case corresponds to the the middle solid curve in figure~\ref{fig:parameter_space_mode3}.

\subsection{Localisation and wave beaming in parabolic metamaterials}

\begin{figure}
\centering
\begin{subfigure}[c]{0.46\linewidth}
\includegraphics[trim = 235 55 160 35, clip,width=\textwidth]{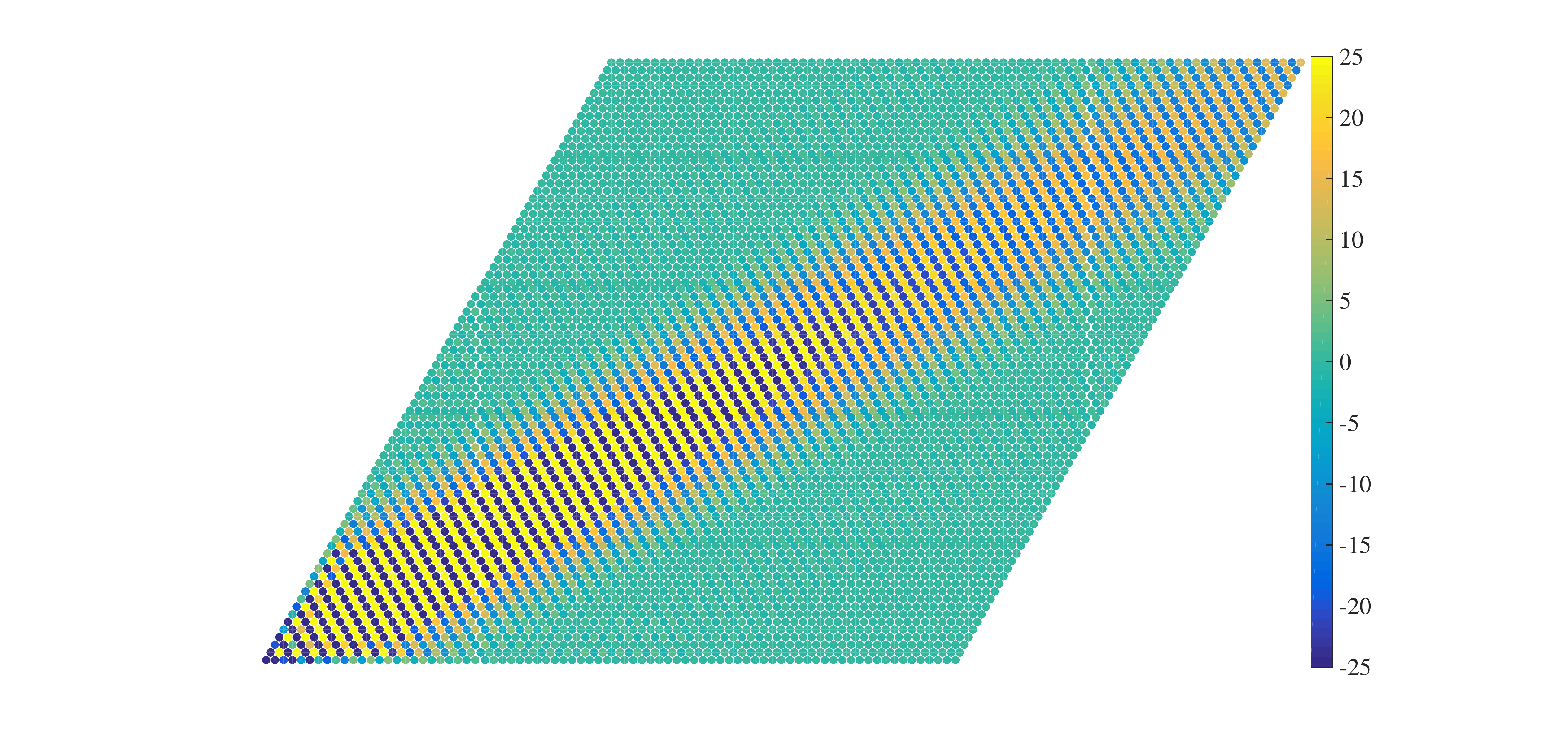}
\caption{\label{fig:gf-mode1}
$u_1(\vec{m})$ for $\beta = \beta^{(2)}_{(11,1)} = 7/18$.}
\end{subfigure}
\begin{subfigure}[c]{0.46\linewidth}
\includegraphics[trim = 235 55 160 35, clip,width=\textwidth]{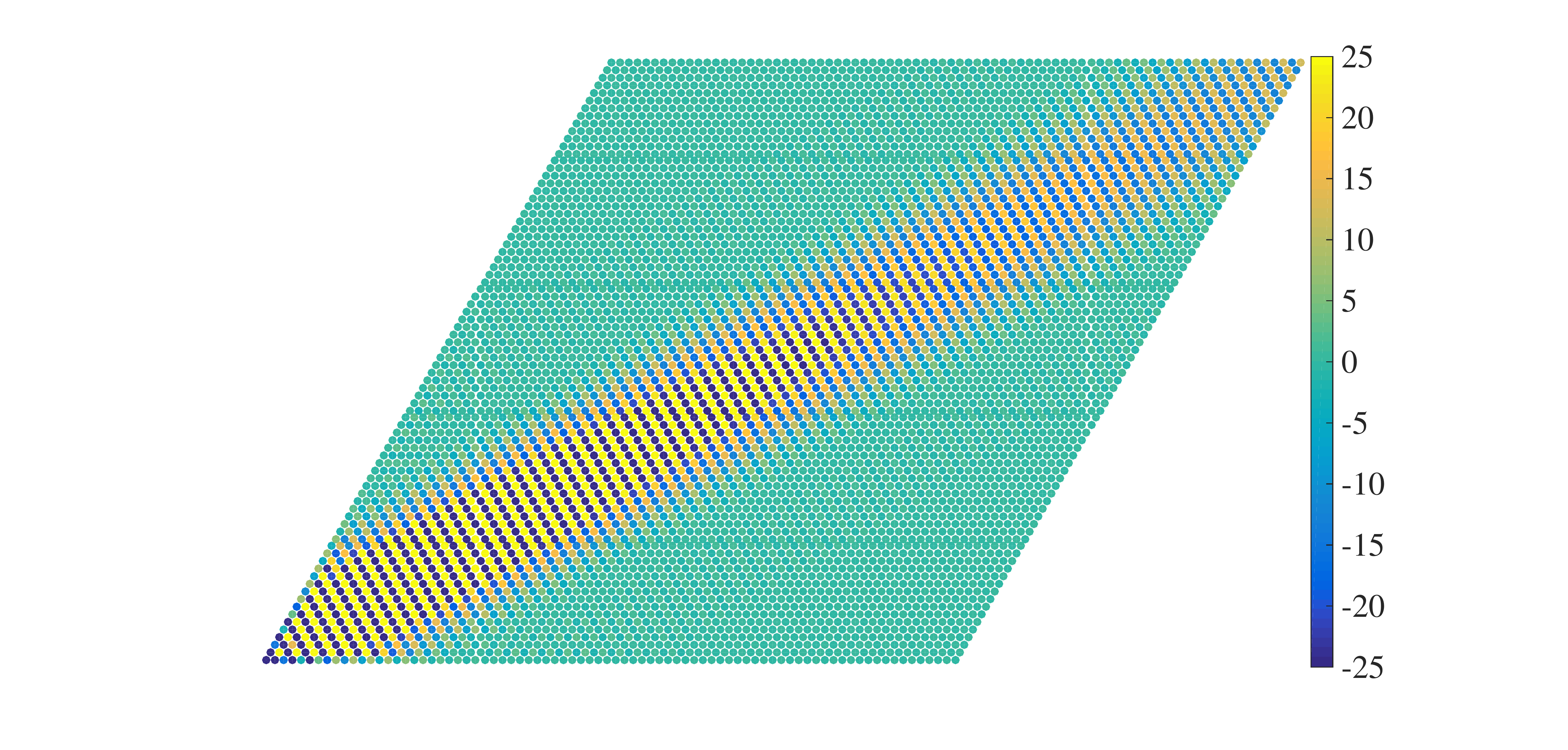}
\caption{\label{fig:gf-mode2}
$u_2(\vec{m})$ for $J=1$ \& $\beta = \beta^{(3)}_{(11,1)} \approx 0.0701$}
\end{subfigure}
\caption{\label{fig:gf}
The highly localised Gaussian beam created by exciting the lattice at the point $\vec{m} = \vec{0}$ in the bottom lefthand corner of the figure.
Here, the excitation frequency corresponds to the dynamic regime in which the lattice behaves as a parabolic metamaterial and the forcing is chosen to be parallel to the corresponding eigenvector.
The left panel shows the first component of the field for the second mode with a time-harmonic forcing of amplitude $\vec{F} = [1,0,0]^\mathrm{T}$, whilst the right panel shows the second component of the field for the third mode with a forcing of amplitude $\vec{F} = [1,0,0]^\mathrm{T}$.
We emphasise that the resulting wave pattern remains essentially unchanged regardless of the chosen orientation of forcing.
The remaining components are vanishingly small away from the excitation point and thus are not shown here.}
\end{figure}

For the purposes of illustration, we now consider the forced problem where the infinite lattice is excited at the origin by a time-harmonic force.
In this case, the equations of motion are
\[
\omega^2\mathsf{M}\vec{u}(\vec{m}) = \sum_{\vec{p}\in\mathcal{N}} \mathsf{C}(\vec{p}) \vec{u}(\vec{m}+\vec{p}) + \vec{F}\delta(\vec{m}),
\]
where $\vec{F}$ is the forcing amplitude and $\delta$ is the discrete Dirac delta.
The response of the lattice can be characterised by the Green's function, which we express in the form of a double Fourier integral
\begin{equation}
\vec{u}(\vec{m}) = \frac{1}{\|\mathcal{R}\|}\iint\limits_\mathcal{R} e^{-i\vec{k}\cdot\mathsf{T}\vec{m}}\left[\sigma(\omega,\vec{k})\right]^{-1}\vec{F}\,\mathrm{d}\vec{k},
\label{eq:gf}
\end{equation}
where $\mathcal{R} = (-2\pi,\pi)\times(-2\pi/\sqrt{3},2\pi/\sqrt{3})$ and $ \sigma(\omega,\vec{k}) = \omega^2\mathsf{M} - \sum_{\vec{p}\in\mathcal{N}} \mathsf{C}(\vec{p})e^{i\vec{k}\cdot\mathsf{T}\vec{p}}$.
Here we are concerned with frequencies that lie within the stop band for which $\det\sigma(\omega,\vec{k}) \neq 0$ and the integrand is well behaved.
In general, equations of the form~\eqref{eq:gf}, cannot be evaluated in closed form, although for lower dimensions or simple configurations some progress can be made (see, for example,~\cite{movchan2007band,colquitt2013localization}).
However, the Green's function~\eqref{eq:gf} is amenable to numerical evaluation, particularly outside the pass band.
This is the approach that we take here where we directly evaluate the double integral~\eqref{eq:gf} using iterative quadrature.

Figure~\ref{fig:gf} shows two configurations where the material parameters have been tuned to achieve an effective parabolic metamaterial.
A point force generates ``beam-like'' fields, where the energy is highly localised within a narrow beam; this effect is due to the parabolic nature of the lattice.
The lattice structure is forced by a point load in the bottom left hand corner and a single unidirectional beam is emitted.
The beam consists of plane waves propagating along the direction defined by the effective parabolic PDE and modulated by a Gaussian-like beam.
In each case we show only the dominant component of the displacement; the omitted components are small at the forcing point and decay rapidly away from the loading.

Figure~\ref{fig:gf-mode1} corresponds to the second mode where we choose $\beta = \beta^{(2)}_{11} = 7/18$.
As remarked earlier, this configuration has the interesting feature that it is parabolic regardless of the chosen value of $J$, meaning that only the stiffness of the lattice links needs to be tuned and the mass of the lattice nodes can be chosen arbitrarily.
This feature makes the practical implementation of parabolic metamaterials more straightforward.
The second panel, figure~\ref{fig:gf-mode2}, corresponds to $\beta = \beta^{(3)}_{(11,1)}$ and $J=1$.
The lattices were excited by a time-harmonic force in the direction parallel to the corresponding eigemodes: $\vec{F} = [1,0,0]^\mathrm{T}$ for figure~\ref{fig:gf-mode1} and $\vec{F} = [0,1,0]^\mathrm{T}$ for figure~\ref{fig:gf-mode2}.
It is interesting to note, however, that the resulting wave pattern remains essentially unchanged regardless of the chosen orientation of forcing; the highly localised unidirectional beam is still present, but the dominant component of displacement may change.

We remark that the numerical evaluation of~\eqref{eq:gf}, although straightforward, is computationally intensive.
Indeed, the computations shown in~\ref{fig:gf} consists of an $80\times80$ grid of nodes at each of which three double integrals corresponding to each component of~\eqref{eq:gf} is evaluated.
The computation was parallelised and implemented in MATLAB.
Although we have not carried out an explicit analysis of the computational cost, it is illuminating to consider that each of the two computations shown in figure~\ref{fig:gf} required approximately 24 hours to complete using 8 dedicated physical cores running at 2.5GHz.
This emphasises the usefulness of the elegant closed form asymptotic solutions presented in this paper.

\section{Concluding remarks}
\label{S4}

In conclusion we emphasise that, for an elastic triangular lattice, the flexural rigidities of the elastic ligaments can be neglected and the longitudinal stiffness plays the dominant role in the determination of the effective material properties.
Such an approximation is fully applicable to static problems.
However, we have demonstrated that the flexural rigidities of the elastic beams within a triangular lattice play an important role in higher-frequency regime linked to description of a new class of metamaterials, which we refer to as `parabolic metamaterials'.

The notion of `hyperbolic metamaterials' is well-established (see, for example, \cite{poddubny2013hyperbolic}) and can be associated with important features of dynamic anisotropy of periodic multi-scale structured media (see~\cite{colquitt2012dynamic,ayzenberg2008resonant,osharovich2010wave,movchan2014resonant,Ruzzene2003,langley1996response,langley1997response}, among others).
However the new class of parabolic metamaterials, identified for the first time in the present paper, is characterised by parabolic effective equations on the long scale, and hence a forced input would produce a highly localised unidirectional Gaussian-like beam in the appropriate dynamic regime.
The notion of a `Dirac bridge' is introduced to describe such dynamic regimes.
These bridges are sections of the dispersion surfaces that connect adjacent Dirac cones.
At the centre of each Dirac bridge is a stationary point associated with a resonance of the system; the dispersion surfaces are locally parabolic in the neighbourhood of these resonances which are associated with the described parabolic behaviour.

The high frequency homogenisation method employed in the present paper leads to explicit formulae for the coefficients of the envelope equations derived here.
It has also been demonstrated how mechanical parameters of the elastic ligaments, such as the moment of inertia of the cross section as well as the ratio of elastic stiffnesses, influence the macroscopic of the metamaterial.
It is interesting to note that, for the triangular lattice, these Dirac bridges only appear when the flexural interaction of the thin elastic ligaments is considered; this is despite the common convention~\cite{deshpande2001foam} to ignore bending moments for so-called `stretch-dominated structures'.
Whilst this may be valid in the classical long wave regime, we have demonstrated here that such interaction play a vital role in the dynamic regime and can lead to novel and exciting effects.

Finally, this analytical study opens perspectives for dynamic analysis of defects and exponential boundary layers at the edges of large structured clusters.  

\section*{Acknowledgements}
A.B.M and N.V.M. gratefully acknowledge the financial support of the EPSRC through programme grant EP/L024926/1.
D.J.C. thanks the EPSRC for support through research grant EP/J009636/1.

\bibliographystyle{qjmam}
\bibliography{references}

\begin{appendices}

\section{The stiffness matrices}
\label{app:matrices}
Although the stiffness matrices $\mathsf{C}(\vec{p})$ can be deduced from those in~\cite[Eq. 9]{colquitt2012dynamic} by taking the limit $\varrho\to0$, we provide them here for convenience
\[
\mathsf{C}(0,0) =
\begin{bmatrix}
3(1+6\beta) & 0 & 0 \\
0 & 3(1+6\beta) & 0 \\
0 & 0 & 12\beta
\end{bmatrix},
\quad
\mathsf{C}(1,0) =  
\begin{bmatrix}
-1 & 0 & 0 \\
0 & -6\beta & 3\beta \\
0 & -3\beta & \beta
\end{bmatrix},
\]
\[
\mathsf{C}(0,1) = -
\frac{1}{4}\begin{bmatrix}
 18\beta + 1 & \sqrt{3}(1-6\beta) & 6\sqrt{3}\beta \\
 \sqrt{3}(1-6\beta) & 6\beta+3 & -6\beta \\
 -6\sqrt{3} \beta & 6\beta & -\beta  \\
\end{bmatrix},
\]
\[
\mathsf{C}(-1,1) = -
\frac{1}{4}\begin{bmatrix}
 18\beta + 1 & \sqrt{3}(6\beta-1) & 6\sqrt{3}\beta \\
 \sqrt{3}(6\beta-1) & 6\beta+3 & 6\beta \\
 -6\sqrt{3} \beta & -6\beta & -\beta  \\
\end{bmatrix},
\]
where stiffness matrices have the property $\mathsf{C}(\vec{p}) = [\mathsf{C}(-\vec{p})]^\dagger$.
\end{appendices}

\end{document}